\documentclass{epl}
\usepackage{graphicx}

\title{Effective phonons in anharmonic lattices: anomalous vs normal heat conduction}
\author{Nianbei Li\inst{1} \and Peiqing Tong\inst{1,3} \and Baowen Li\inst{1,2,4}
\thanks{E-mail: \email{phylibw@nus.edu.sg}}}

\institute{
  \inst{1} Department of Physics and Center for
Computational Science and Engineering, National University of
Singapore, Singapore 117542,
 Republic of Singapore\\
  \inst{2} Laboratory of Modern Acoustics and Institute of Acoustics,
 Nanjing University, 210093, P R China\\
  \inst{3} Department of Physics, Nanjing Normal University,
 Nanjing, Jiangsu
 210097, PR China\\
  \inst{4} NUS Graduate School for Integrative Sciences and Engineering,
Singapore 117597, Republic of Singapore
}

\begin{document}

\maketitle

\begin{abstract}
We study heat conduction in one dimensional (1D) anharmonic
lattices analytically and numerically by using an effective phonon
theory. It is found that every effective phonon mode oscillates
quasi-periodically. By weighting the power spectrum of the total
heat flux in the Debye formula, we obtain a unified formalism that
can explain anomalous heat conduction in momentum conserved
lattices without on-site potential and normal heat conduction in
lattices with on-site potential. Our results agree very well with
numerical ones for existing models such as the Fermi-Pasta-Ulam
model, the Frenkel-Kontorova model and the $\phi^4$ model etc.

\end{abstract}

Recent years has witnessed increasing studies on heat conduction
in one dimensional (1D) anharmonic (nonlinear)
lattices\cite{Review}. On the one hand, people would like to know
whether or not the Fourier's law of heat conduction for bulk
material is still valid in 1D systems. This is a fundamental
question in non-equilibrium statistical mechanics. In fact, it is
not trivial at all as a rigorous proof is still not possible. On
the other hand, the fast development of nano technology makes it
possible to fabricate 1D or quasi 1D systems such as nanowire
and/or nanotube etc and to measure its transport properties. To
understand heat conduction behavior in such systems is of great
interest in heat control and management at nanoscale. Numerically,
an anomalous heat conduction  - heat conductivity diverges with
system size- has been observed in momentum conserved systems
without on-site potential such as the Fermi-Pasta-Ulam (FPU)
lattice\cite{FPUheat}, and a normal heat conduction has been found
in the systems with on-site potential like the Frenkel-Kontorova
model\cite{HLZ98} and the $\phi^4$ model\cite{HLZ00,Aoki1}.
Unfortunately, up to now a general theory to predict the heat
conduction behavior in a 1D system is still lacking.

In this Letter, we investigate analytically and numerically the
physical mechanism leading to the anomalous and the normal heat
conduction in 1D anharmonic (nonlinear) lattices from an {\it
effective phonon theory}. The theory is based on the ergodic
hypothesis (equipartition theorem). As it will be seen that, our
analytical result can explain the anomalous and normal heat
conduction observed numerically in different models.

We consider a 1D anharmonic (nonlinear) lattice with the
Hamiltonian
\begin{equation}
H=\sum^{N}_{i=1}\left[\frac{1}{2}m_i\dot{x}^2_i+V(\delta
x_{i,i+1})+U(x_i)\right] \label{eq:Ham}
\end{equation}
with periodic boundary condition $x_1= x_{N+1}$. Here we chose
mass $m_i=1$ for all lattice. $\dot{x}_i=dx_i/dt$.  $\delta
x_{i,i+1}=x_i-x_{i+1}$. Without loss of generality, we write the
inter particle potential $V(\delta x_{i,i+1})$ and the on-site
potential $U(x_i)$ as
\begin{equation}
 V(\delta
x_{i,i+1})=\sum^{\infty}_{s=2}g_{s}\frac{(\delta
x_{i,i+1})^{s}}{s},\,\,\,U(x_i)=\sum^{\infty}_{s=2}\sigma_{s}\frac{x_i^{s}}{s},
\end{equation}
respectively. The canonical transformation which diagonalizes the
harmonic Hamiltonian is ${\bf X}={\bf BQ}$, where $X_i=(x_i\,
\mbox{or}\, \dot{x}_i)$, $Q_k=(q_k\,\mbox{or}\, p_k)$, and
$B_{ik}$ are\cite{Alabiso1}
$$
B_{ik} = \left\{
\begin{array}{ll}
\sqrt{\frac{2}{N}}G_k\cos{\frac{2i\pi(k-1)}{N}} & k=1,...,\left[\frac{N}{2}\right]+1\\
\sqrt{\frac{2}{N}}G_k\sin{\frac{2i\pi(N-k+1)}{N}} &
k=\left[\frac{N}{2}\right]+2,...,N
\end{array} \right.
$$
where $\left[\frac{N}{2}\right]$ is the integer part of
$\frac{N}{2}$, and $G_{k}=1/\sqrt{2}$ for $k=1$ and $k=N/2+1$ if
$N$ is even, otherwise $G_{k}=1$. The spectrum of harmonic lattice
is $\omega_k=2\sin{(k-1)\pi/N}$.

Under the {\it ergodic hypothesis}, the system obeys the
generalized equipartition theorem\cite{Segal}
$k_{B}T=\left<q_{k}\frac{\partial{H}}{\partial{q_k}}\right>$, here
$\left<\cdot\right>$ denotes the canonical ensemble average. The
force in $k$ space has two parts
$$
-F_k=\frac{\partial{H}}{\partial{q_k}}=\sum^{N}_{i=1}\sum^{\infty}_{s=2}
\left(\omega_{k}g_{s}(\delta x_{i,i+1})^{s-1}\gamma_{ik}
+\sigma_{s}x_{i}^{s-1}B_{ik}\right),$$
 where the new matrix
$\gamma_{ik}$ is defined as in Ref.\cite{Alabiso1},
$\gamma_{ik}=0$ for $k=1$ and
$\gamma_{ik}=(1/\omega_k)(B_{ik}-B_{i+1k})$ otherwise.
$\gamma_{ik}$ satisfy $ \sum^{N}_{i=1}\delta
x_{i,i+1}\gamma_{ik}=\omega_{k}q_k,\,\,\,
\sum_{k=2}^{N}\gamma_{ik}\omega_{k}q_{k}=\delta x_{i,i+1}. $ The
generalized equipartition theorem becomes
\begin{eqnarray}\label{equi}
k_{B}T
&=&\sum^{N}_{i=1}\sum^{\infty}_{s=2}\left(\omega_{k}g_{s}\langle(\delta
x_{i,i+1})^{s-1}q_k\rangle\gamma_{ik}
+\sigma_{s}\langle x_{i}^{s-1}q_{k}\rangle B_{ik}\right)\nonumber\\
&\approx &\sum^{\infty}_{s=2}\left[g_{s}
\frac{\langle\sum^{N}_{i=1}(\delta
x_{i,i+1})^{s}\rangle}{\langle\sum^{N}_{i=1}(\delta
x_{i,i+1})^{2}\rangle} \omega^2_k
+\sigma_{s}\frac{\langle\sum^{N}_{i=1}x_{i}^{s}\rangle}{\langle
\sum^{N}_{i=1}x_{i}^{2}\rangle}
\right]\langle q^2_{k}\rangle\nonumber\\
&\equiv&\alpha(\omega^2_{k}+\gamma)\langle q^2_{k}\rangle
\end{eqnarray}
where
$$\alpha=\frac{\sum^{\infty}_{s=2}g_{s}
\langle\sum^{N}_{i=1}(\delta
x_{i,i+1})^{s}\rangle}{\langle\sum^{N}_{i=1}(\delta
x_{i,i+1})^{2}\rangle}, $$ and
$$\gamma=\frac{1}{\alpha}\frac{\sum_{s=2}^{\infty}\sigma_{s}\langle\sum^{N}_{i=1}x_{i}^{s}
\rangle}{\langle \sum^{N}_{i=1}x_{i}^{2}\rangle}.$$

In analogy with an Harmonic lattice where
$k_{B}T=\omega^2_{k}\langle q^2_{k}\rangle$,  we define effective
phonons in 1D anharmonic lattices. The frequencies of the
effective phonon are
\begin{equation}
\hat{\omega}^2_{k}=\alpha(\omega^2_{k}+\gamma),
\end{equation}
here $k$ is wave vector with replacement $k\rightarrow
2\pi(k-1)/N$ and $\omega_{k}=2\sin{\frac{k}{2}}$. The
corresponding velocities are
\begin{equation}
v_k=\frac{\partial{\hat{\omega}_k}}{\partial{k}}
=\frac{\sqrt{\alpha}\,\,\omega_{k}}{\sqrt{\omega^2_{k}+\gamma}}\cos{\frac{k}{2}}.
\end{equation}

 It has been found
\cite{Alabiso1,Alabiso2} numerically  that the approximation in
Eq. (\ref{equi}) is feasible for anharmonic lattices without
on-site potential such as the FPU-$\beta$ model ($U(x)=0$,
$V(x)=x^2/2+\beta x^4/4$, we take $\beta=1$ in this paper) and
$H_4$ model ($U(x)=0$, $V(x)=x^4/4$.), $\alpha$ is found to be
independent of mode $k$ and lattice length.

\begin{figure}
\twofigures{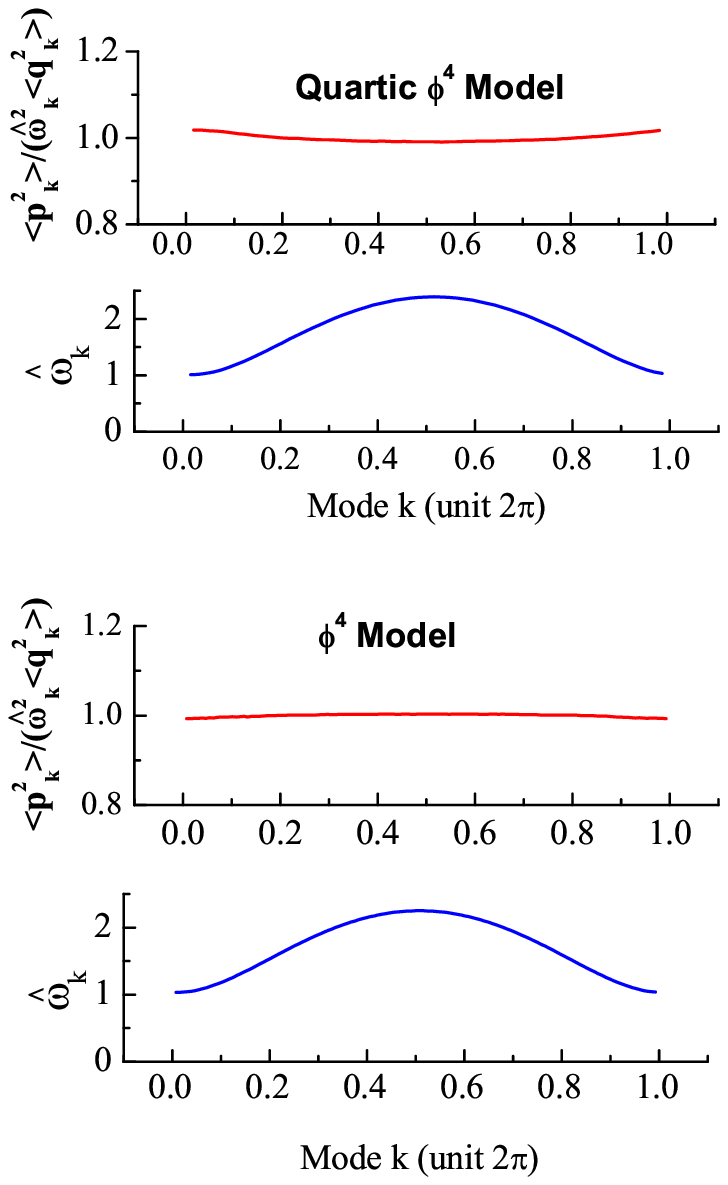}{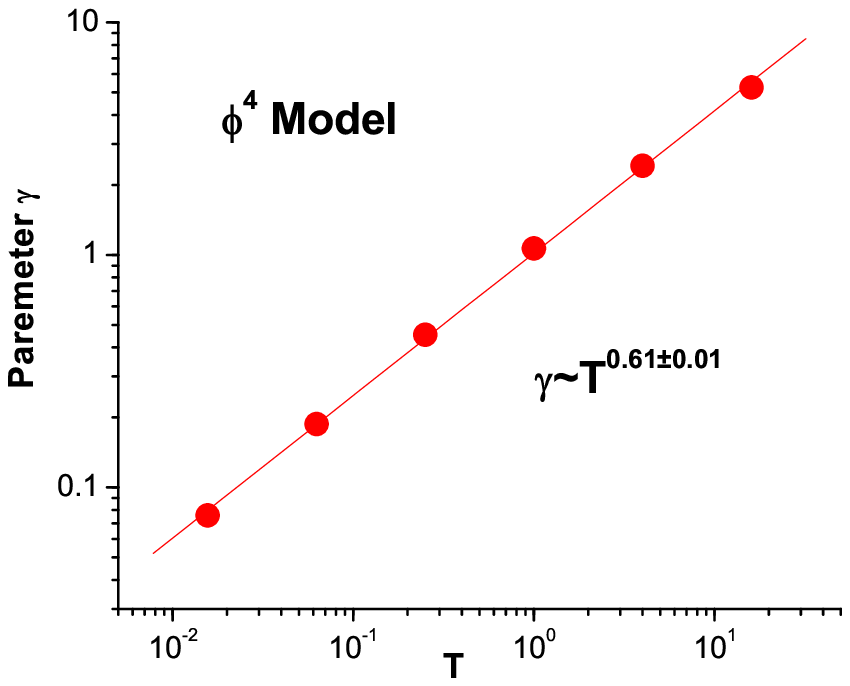} \caption{The energy
ratio, $\langle p^2_k\rangle/\hat{\omega}^2_{k}\langle
q^2_k\rangle$, and dispersion relation $\hat{\omega}_k$ versus $k$
for the $\phi^4$ model (lower panel) and the quartic $\phi^4$
model (upper panel). $N=128$ for the $\phi^4$ model, and $N=64$
for the quartic $\phi^4$ model. Temperature is set to $T=1$.}
\label{fig:phi4}\caption{Parameter $\gamma$ versus temperature $T$
for the $\phi^4$ model.} \label{fig:gammat}
\end{figure}

In order to check if Eq. (\ref{equi}) is suitable for the
anharmonic chains with on-site potential, we study numerically
$\gamma$ for the $\phi^4$ model ($U(x)= x^2/2, V(x)=x^4/4$) and
the quartic $\phi^4$ model ($U(x)=x^4/4, V(x)=x^4/4$). We find
that $\gamma$ is independent of $k$ and the chain length provided
the chain is long enough. For example, $\gamma\simeq 1.065$ for
the $\phi^4$ model with chain length $N\geq 64$ at temperature
$T=1$. In Fig.\ref{fig:phi4}, we show $\langle
p^2_k\rangle/(\hat{\omega}^2_k\langle q^2_k\rangle)$ and
dispersion relation $\hat{\omega}_k$ for quartic $\phi^4$ (upper
panel) and $\phi^4$ (lower panel) models. $\langle
p^2_k\rangle/(\hat{\omega}^2_k\langle q^2_k\rangle)$ is very close
to one for all $k$ in both models. We also find that $\gamma$
increases with temperature as $\gamma \sim T^{0.61\pm0.01}$ (see,
Fig. ~\ref{fig:gammat}). Therefore, we can conclude that the
approximation is good for both anharmonic chains with and without
on-site potential under ergodic hypothesis.

It should be pointed out that the approximation in Eq.
(\ref{equi}) is a kind of mean-field approximation. It is based on
our numerical observation as a rigorous proof is not possible,.
The phonon-phonon interaction is implicitly contained in the two
coefficients $\alpha$ and $\gamma$.

\begin{figure}
\twofigures{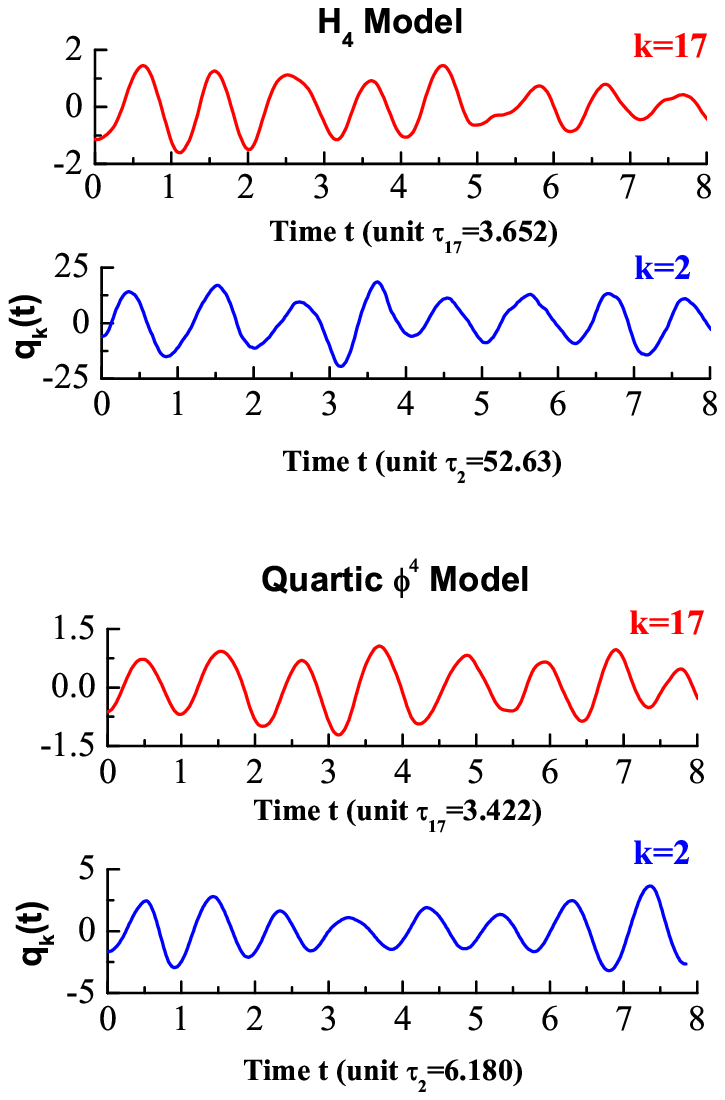}{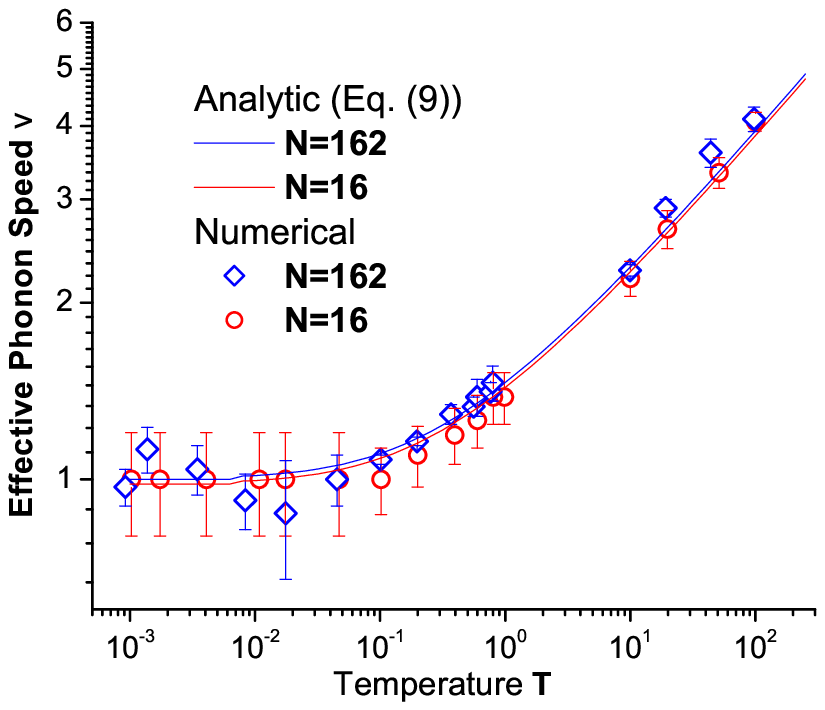}\caption{Time behaviors of
$q_k(t)$ of the $H_4$ model and the quartic $\phi^4$ model. $N=64$
and $T=1$ for both models. $\tau_k=2\pi/\hat{\omega}_k$.}
\label{fig:qkt} \caption{The effective phonon speed in FPU-$\beta$
model vs temperature. The numerical results come from the lower
panel of Fig. 2 in Ref. \cite{Aoki2}} \label{fig:fpuspeed}
\end{figure}

The heat conductivity can be derived from the Debye formula
$$\kappa=\sum_{k}c_{k}v^2_k\tau_k,$$
where $c_k$, $v_k$, and $\tau_k$ are specific heat, phonon
velocity and phonon relaxation time of mode $k$, respectively.
Generally, the contribution from phonons of different frequencies
should in principle be weighted but this is not reflected in the
Debye formula. Since the heat conductivity is a transport
coefficient of heat energy which manifest itself by heat flux,
this weight factor must be the power spectrum of total heat flux.
Thus we rewrite the Debye formula for 1D anharmonic lattices as,
$$
\kappa=\sum_{k}P_{k}c_{k}v^2_k\tau_k,\,\,0< k\leq 2\pi, $$ in
which $P_{k}$ is the normalized power spectrum of total heat flux.
Moreover, the Debye formula does not give an explicit expression
of phonon relaxation time $\tau_k$. From our numerical
calculations, we find that each mode oscillates quasi-periodically
for both lattices with and without on-site potential (see
Fig.\ref{fig:qkt}). It is therefore reasonable to assume that the
phonon relaxation time is proportional to the quasi-period of each
mode, {\sl i.e.}, $\tau_k= \lambda \frac{2\pi}{\hat{\omega}_k}$,
where prefactor $\lambda$ is only temperature dependent and will
be discussed in another paper.

To deal with the size dependence of the heat conductivity, it is
more convenient to consider the mean-free-path rather than the
relaxation time. The mean-free-path of the effective phonons is
defined by
\begin{equation}
l_k=v_k
\tau_k=2\pi\lambda\frac{\omega_k}{\omega^2_k+\gamma}\cos{\frac{k}{2}}
\end{equation}
For systems without on-site potential where $\gamma=0$, $l_k$
reduces to
$$l_k=\frac{2\pi\lambda}{\omega_k}\cos{\frac{k}{2}}.$$
In the long wave-length limit, $k\rightarrow0$, the mean-free-path
$l_k\propto 1/k$, becomes divergent. However, for systems with
on-site potential where $\gamma>0$, the mean-free-path of any
effective phonon is finite. This difference results in different
heat conduction behaviors in systems without on-site potential and
systems with on-site potential.

The  modified Debye formula of thermal conductivity can be
expressed in a continuous form in the thermodynamical limit:
\begin{equation}\label{kappa}
\kappa =\frac{N}{2\pi}\int^{2\pi}_{0}P(k)c_k v_k l_k dk
=c\lambda\sqrt{\alpha}\int^{2\pi}_{0}P(k) \frac{\omega^2_{k}}
{\left(\omega^2_{k}+\gamma\right)^{3/2}}\cos^{2}{\frac{k}{2}} dk
\end{equation}
where $c=\sum_k c_k$. By definition, $P(k)dk=P(\omega)d\omega$,
and $P(\omega)$ is the Fourier transform of auto-correlation
function of total heat flux $J(t)$.

Eq. (\ref{kappa})is the main analytical result of this Letter.
Whether the heat conduction in a 1D system is normal or anomalous
depends on whether the integral of Eq. (\ref{kappa}) is finite or
infinite. For systems with on-site potential such as the FK model
and the $\phi^4$ model where $\gamma>0$, all integrands except the
normalized power spectrum $P(k)$ are finite. Since $P(k)$ should
be normalized over the phonon spectrum $\int^{2\pi}_{0}P(k)dk=1$,
the integral of Eq. (\ref{kappa}) for systems with on-site
potential is always finite. Thus even without knowing exact
knowledge of $P(k)$, we can predict that heat conduction of
systems with on-site potential obeys Fourier's Law. However for
systems without on-site potential, $\gamma=0$ which implicitly
means the momentum conservation. The integral of Eq. (\ref{kappa})
reduces to $\int^{2\pi}_{0}P(k)
\frac{\cos^{2}{\frac{k}{2}}}{\omega_{k}}dk$. In the long
wave-length limit,  $\omega_{k} \approx k$, the integral has a
singularity at $k\rightarrow 0$. This singularity originates from
the infinite mean-free-path of effective phonon of the long
wave-length limit and is proportional to $1/k$. Since the
effective phonon of long wave-length is the most dominant part for
heat transfer, the systems without on-site potential will exhibit
an anomalous heat conduction. In the following we shall apply
above theory to two typical classes of Hamiltonian systems, the
FPU model, a representative model without on-site potential, and
the $\phi^4$ model, a representative one with on-site potential.
As shall be seen soon that our theory gives predictions in good
agreements with numerical simulations.

{\bf Systems without On-Site Potential}. In the Hamiltonian
(\ref{eq:Ham}), $U(x)=0$ means momentum conservation, and
 $\gamma=0$ in Eq.(\ref{equi}). Here we focus on the FPU-$\beta$ model.
 The details of renormalized frequencies of
effective phonons will be discussed as well as the effective
phonon speed and the heat conductivity with respect to lattice length.

In the FPU-$\beta$ model, $V(x)=x^2/2+ x^4/4$ and $U(x)=0$.
$\alpha$ is determined only by temperature\cite{Alabiso1,Lepri3}.
In this model, $\alpha$ has a simple analytic expression
\begin{equation}
\alpha=1+\frac{\left<\sum^{N}_{i=1}\left(\delta{x}_{i,i+1}\right)^4\right>}
{\left<\sum^{N}_{i=1}\left(\delta{x}_{i,i+1}\right)^2\right>}
=1+\frac{\int^{\infty}_{-\infty}\phi^{4}e^{-V(\phi)/T}d\phi}
{\int^{\infty}_{-\infty}\phi^{2}e^{-V(\phi)/T}d\phi}
\end{equation}
this equation is equivalent to the Eq. 11 of Ref. \cite{Lepri3}.

If the  effective phonons of long wave-length are dominant in heat
transfer, which is true for a system of large size, the speed of
energy transport can be approximated by the speed of effective
phonon with the longest wave-length
\begin{equation}
v=\sqrt{\alpha}\cos{\frac{\pi}{N}}.
\label{Eq:PhononSpeed}
\end{equation}
In Fig. \ref{fig:fpuspeed} we draw above analytical $v$ versus
temperature for two different system sizes and compare them with
numerical ones from Aoki and Kuznezov\cite{Aoki2}. The agreement
between our analytical results (\ref{Eq:PhononSpeed}) and
numerical ones are very good in a very wide range of temperature (
five orders of magnitudes.)

In the thermodynamic limit $N\rightarrow\infty$, the size
dependence of heat conductivity only depends on the integral
\begin{equation}
\kappa\propto\int^{2\pi}_{0}P(k)\frac{\cos^{2}
{\frac{k}{2}}}{\omega_{k}}dk.
\end{equation}
If the  effective phonons of long wave-length dominate the heat
transfer, and $P(k)\propto k^{-\delta}$ ($\delta>0$)
asymptotically in long wave-length limit, then, the heat
conductivity is a system size dependent quantity, $\kappa\propto
k^{-\delta} \propto N^{\delta}$. Numerical calculation of power
spectrum of total heat flux in FPU-$\beta$ model has shown that
$\delta\approx0.37\sim0.4$\cite{FPUheat}. This divergent behavior
of thermal conductivity has been observed in the FPU model by
different groups \cite{FPUheat,HLZ00,Aoki2}.

{\bf Systems with On-Site Potential} Things turn out to be
different for systems with on-site potential. Having an on-site
potential means that $\gamma>0$ and the momentum conservation is
broken. A phonon band gap appears around zero frequency. Under the
ergodic assumption, heat conduction in such systems obeys the
Fourier's law. For the sake of simplicity, we only discuss the
$\phi^4$ model.

The $\phi^4$ model has a quadratic inter particle potential, $
V(x)=x^2/2$ and a quartic external potential $U(x)=x^4/4$.
Therefore, we have $\alpha=1$ and
$\gamma=\langle\sum^{N}_{i=1}x_i^{4}
\rangle/\langle\sum^{N}_{i=1}x_i^{2}\rangle$. The ensemble average
has no simple analytic expression. At temperature $T=1$, the
numerical value $\gamma\approx1.065$ is found to be independent of
length at least for $N\geq64$. The dispersion relation,
$\hat{\omega}_k=\sqrt{4\sin^2{\frac{k}{2}}+1.065}$, is shown in
Figure \ref{fig:phi4}.

The mean-free-path $l_k$ depends on temperature via parameter
$\gamma$ and $\lambda$, $ l_k=v_k\tau_k=2\lambda\pi\sin k/(4\sin^2
\frac{k}{2}+\gamma)$. In the low temperature limit, the $\phi^4$
model reduces to the harmonic model where $\lambda$ is infinity.
Therefore, $\lambda$ should decrease with temperature. From Fig.
\ref{fig:gammat}, we can see that $\gamma$ increases monotonically
with temperature. As a result, the mean-free-path decreases as
temperature is increased. Since the heat conductivity is
independent of $N$ when $N$ is larger than the mean-free-path. The
higher the temperature, the smaller the size effect for the
$\phi^4$ model as observed by Aoki and Kuznezov\cite{Aoki2}.

In summary, from an effective theory, we have derived an analytic
formula for heat conductivity in 1D nonlinear lattices, Eq.
(\ref{kappa}). We find that the phonon-phonon interaction due to
the anharmonicity (nonlinearity) can be written as an effective
harmonic one in terms of ensemble average, so we can attribute the
heat transfer to the effective phonons which can be treated in the
same way as phonons in the Harmonic lattice. The difference
between system without on-site potential and system with on-site
potential lies in their renormalized frequencies. For systems
without on-site potential, there exists zero frequency mode with
infinite mean-free-path which is the physical mechanism for
anomalous heat conduction. For systems with on-site potential, a
phonon band gap near zero frequency appears leading to the normal
heat conduction, i.e. heat conduction obeys the Fourier's Law.

\acknowledgments We would like to thank K. Aoki and D. Kusnezov
for providing the data of Fig. 2 in Ref\cite{Aoki2}, and L Wang
and P Hanggi for helpful discussions and comments. This work is
supported in part by a FRG grant of NUS and the DSTA under Project
Agreement POD0410553. PQT is supported in part by the National
Science Foundation of China (Grants nos. 90203009 and 10175035)

\end{document}